\documentclass[pra,twocolumn,showpacs,superscriptaddress]{revtex4}

\usepackage{graphicx}
\usepackage{amsmath}
\usepackage{amssymb}

\begin{document}
\title{Entanglement evaluation of non-Gaussian states
generated by photon subtraction from squeezed states}
\author{Akira Kitagawa}
\email{kitagawa@nict.go.jp}
\affiliation{National Institute of Information and Communications
Technology (NICT) 4-2-1 Nukui-Kita, Koganei, Tokyo 184-8795 Japan}
\affiliation{Core Research for Evolutional Science and Technology
(CREST),
Japan Science and Technology Agency \\
1-9-9 Yaesu, Chuoh, Tokyo 103-0028 Japan}
\author{Masahiro Takeoka}
\affiliation{National Institute of Information and Communications
Technology (NICT) 4-2-1 Nukui-Kita, Koganei, Tokyo 184-8795 Japan}
\affiliation{Core Research for Evolutional Science and Technology
(CREST),
Japan Science and Technology Agency \\
1-9-9 Yaesu, Chuoh, Tokyo 103-0028 Japan}
\author{Masahide Sasaki}
\affiliation{National Institute of
Information and Communications Technology (NICT) 4-2-1 Nukui-Kita,
Koganei, Tokyo 184-8795 Japan} \affiliation{Core Research for
Evolutional Science and Technology (CREST),
Japan Science and Technology Agency \\
1-9-9 Yaesu, Chuoh, Tokyo 103-0028 Japan}
\author{Anthony Chefles}
\affiliation{School of Mathematical Sciences, University
College Dublin, Dublin 4, Ireland}
\date{\today}
\begin{abstract}
\vspace{0.5cm}

We consider the problem of evaluating the entanglement of
non-Gaussian mixed states generated by photon subtraction from
entangled squeezed states. The entanglement measures we use are the
negativity and the logarithmic negativity. These measures possess
the unusual property of being computable with linear algebra
packages even for high-dimensional quantum systems. We numerically
evaluate these measures for the non-Gaussian mixed states which are
generated by photon subtraction with on/off photon detectors. The
results are compared with the behavior of certain operational
measures, namely the teleportation fidelity and the mutual
information in the dense coding scheme. It is found that all of
these results are mutually consistent, in the sense that whenever
the enhancement is seen in terms of the operational measures, the
negativity and the logarithmic negativity are also enhanced.

\end{abstract}

\pacs{03.67.Mn, 03.67.Hk, 42.50.Dv}

\maketitle

\section{Introduction}

Continuous variable (CV) quantum optical systems are
well-established tools for both theoretical and experimental
investigations of quantum information processing (QIP)
\cite{Nielsen,Braunstein03}. The essential resource, entanglement,
can be realized with a \textit{Gaussian} two-mode squeezed vacuum
state. This state is relatively easy to work with theoretically and
is also commonly produced in the laboratory. It has been
successfully applied to implement various important protocols, such
as quantum teleportation
\cite{Braunstein98,Furusawa98,Zhang03,Bowen03}, quantum dense coding
\cite{Ban99,Li02,Mizuno05} and entanglement swapping
\cite{Jia04,Takei05}. These successes are based on well-developed
techniques of optical Gaussian operations. These consist of beam
splitting, phase shifting, squeezing, displacement and homodyne
detection.

However, recent theoretical investigations have shown some limits on
such operations. A prime example is the no-go theorem relating to
the distillation of entanglement shared by distant parties using
only Gaussian local operations and classical communication (LOCC)
\cite{Eisert02,Fiurasek02,Giedke02}. To go beyond this limit, one
should use higher order nonlinear processes, such as the cubic-phase
gate \cite{Gottesman01} or Kerr nonlinearity \cite{Nemoto04}. It is,
however, difficult to implement these nonlinear processes with
presently available materials, which do not have sufficiently high
nonlinearity and suffer from losses. A more practical alternative
method has been developed, which uses nonlinear processes induced by
photon counting on tapped-off beams from squeezed states
\cite{Opatrny00,Cochrane02,Browne03}. When ideal photon number
resolving detectors are used, the input Gaussian state can be
transformed into a non-Gaussian pure state with higher entanglement.
In practice, however, such detectors are not yet suitable for
practical use. The most reliable type of photodetector available at
present is the on/off type photon detector based on avalanche
photodiodes. This device can only distinguish the vacuum (`off')
state from non-vacuum (`on') states. The latter events result in a
non-Gaussian mixed state. Evaluating the entanglement of such a
state is far from trivial.

Previously, the effect of entanglement has been theoretically
analysed based on the figures of merit of concrete protocols, such
as the fidelity of teleportation \cite{Olivares03}, the degree of
violation of Bell-type inequalities
\cite{Nha04,Garcia-Patron04,Olivares04} and the mutual information
of dense coding \cite{Kitagawa05}. In fact, it was shown that the
performance of every protocol was improved, implying that the
entanglement of the non-Gaussian mixed state must be enhanced.
However, these indirect evaluations were dependent upon some
external parameters, which depend upon specific operations differing
from protocol to another, such as the type of input state for the
teleportation fidelity, the choice of the measurement basis for the
Bell-type inequality violation, and the signal power of modulation
for the dense coding.

Quantifying the entanglement of the photon-subtracted squeezed
state, in a way which is independent of particular external
parameters, is our main concern in this paper.   To our knowledge,
this problem has not been previously addressed. For a state
$\hat{\rho}$, an entanglement measure $E(\hat{\rho})$ should satisfy
the following criteria \cite{Vidal00}: (i) $E$ is the non-negative
functional, (ii) $E$ vanishes if the state $\hat{\rho} $ is
separable and (iii) $E$ should not increase \textit{on average}
under LOCC. In general, a quantity satisfying these criteria is
known as an entanglement monotone. Many such quantities have been
proposed, such as the entanglement of formation
\cite{Bennett96-2,Wootters98}, the entanglement cost
\cite{Hayden01}, the distillable entanglement \cite{Bennett96-2},
and the relative entropy of entanglement \cite{Vedral98}. It is,
however, not easy to calculate these measures for generic mixed
states. Recently, however, two entanglement measures which are much
more amenable to evaluation have been proposed. These are the
negativity and the logarithmic negativity \cite{Vidal02}. These
measures are based on the Peres criterion \cite{Peres96}. That is,
they are defined in terms of the eigenvalues of the
partially-transposed density operator. The most distinctive feature
of these entanglement measures is that they are easily computable
numerically with linear algebra packages. Furthermore, the
logarithmic negativity is an additive functional, and it is an upper
bound on the distillable entanglement $E_{\rm D}$ \cite{Vidal02}.

In addition to being an entanglement monotone, it was believed that
an entanglement measure $E$ should also be (downward) convex, i.e.
it should be non-increasing on average under the loss of classical
information by mixing. Convexity is necessary for an entanglement
measure to be bounded from above by the entanglement of formation
$E_{\rm F} $ \cite{Horodecki00}, although the logarithm function is
concave (upward convex). Recently, however, it has been shown by
Plenio {\em et al} \cite{Plenio05,Plenio05-2} that convexity is not
directly related to the physical process of discarding {\em quantum}
information, i.e. discarding subsystems such as local ancillas
during LOCC operations
 and that the logarithmic negativity is
indeed a full entanglement monotone \cite{Plenio05-2}. In view of
these considerations, it is interesting to evaluate the entanglement
of photon-subtracted mixed state in terms of the (logarithmic)
negativities, and to compare them the figures of merit associated
with the aforementioned protocols.  Of particular interest to us is
how the entanglement, and these figures of merit, are enhanced by
the photon-subtraction procedure.

This paper is organized as follows. In Sec. \ref{MING-op}, we
briefly summarize the measurement-induced non-Gaussian operation on
the two-mode squeezed vacuum state and its mathematical description.
In Sec. \ref{monotonicity}, the negativity and the logarithmic
negativity are briefly reviewed, and their monotonicity is
discussed. In Sec. \ref{numerical result}, our numerical methods for
calculating the negativities of the photon-subtracted mixed state
are presented. In Sec. \ref{op_meas}, we review the previous
analyses of the entanglement of such states using protocol-specific
figures of merit.   We compare these results with those relating to
the negativity/logarithmic negativity.  The final section
\ref{discussion} is devoted to discussion and conclusion.

\section{Measurement-induced non-Gaussian operation}
\label{MING-op}

\begin{figure}
\centering
\includegraphics[bb=55 55 345 340, width=.8\linewidth]{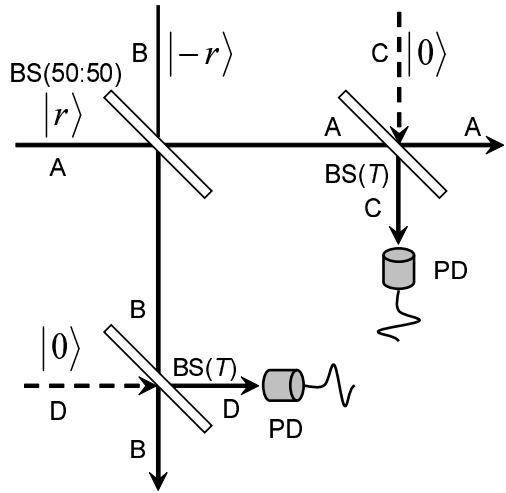}
\caption{\label{2mode-PS} Measurement-induced non-Gaussian operation
on the two-mode squeezed vacuum state. BS, PS are beam splitter,
photon detector, respectively. }
\end{figure}

The schematic of the measurement-induced non-Gaussian operation
on the two-mode squeezed state is shown in Fig. \ref{2mode-PS}.
The primary sources are two identical, single-mode squeezed vacuum
states,
\begin{equation}
|r\rangle _k =\hat{S}_k (r)|0\rangle _k ,
\end{equation}
where $\hat{S}_k (r)$ is the squeezing operator at path $k$,
\begin{equation}
\hat{S}_k (r)=\exp \left[ -\frac{r}{2} (\hat{a}_k^{\dagger 2}
-\hat{a}_k^2 )\right] ,
\end{equation}
and $r$ is the squeezing parameter. These are combined via a
balanced beam splitter to generate the two-mode squeezed vacuum
state,
\begin{eqnarray}
|r^{(2)} \rangle _{\rm AB} &=&\hat{V}_{\rm AB} \left( \frac{\pi }{4} \right)
|r\rangle _{\rm A} |-r\rangle _{\rm B} \nonumber \\
&=&\hat{S}_{\rm AB} (-r)|0\rangle _{\rm AB} \nonumber \\
&=&\sum _{n=0}^\infty \alpha _n |n\rangle _{\rm A} |n\rangle _{\rm
B} , \label{SVS}
\end{eqnarray}
where
\begin{equation}
\hat{V}_{kl} (\theta )=\exp \left[ \theta (\hat{a}_k^\dagger
\hat{a}_l -\hat{a}_k \hat{a}_l^\dagger )\right]
\end{equation}
is the beam splitter operator, and the parameter $\theta $ is
related to the transmittance $T$ as
\begin{equation}
\tan \theta =\sqrt{\frac{1-T}{T} },
\end{equation}
and $\theta =\pi /4$ corresponds to the balanced beam splitter.
The operator $\hat{S}^{(2)} _{kl} (r)$ is the two-mode squeezing operator,
\begin{equation}
\hat{S}^{(2)} _{kl} (r)=\exp \left[ -r(\hat{a}_k^\dagger \hat{a}_l
^\dagger -\hat{a}_k \hat{a}_l )\right] .
\end{equation}
Introducing $\lambda = \tanh r$, the Schmidt coefficients are given
by
\begin{equation}
\alpha _n =\sqrt{1-\lambda^2} \lambda ^n . \label{Schmidt}
\end{equation}

The beam at path C (D) is then tapped off from path A (B)
by a beam splitter of transmittance $T$.
The resulting four-mode state just after the second beam splitters is
\begin{eqnarray}
|\psi \rangle _{\rm ABCD} &=&\hat{V}_{\rm BD} (\theta )\hat{V}_{\rm AC} (\theta )
|r^{(2)} \rangle _{\rm AB} |0\rangle _{\rm CD} \nonumber \\
&=&\sum _n \alpha _n \sum _{i,j=0}^n \xi _{ni} \xi _{nj} |n-i\rangle
_{\rm A} |n-j\rangle _{\rm B} |i\rangle _{\rm C} |j\rangle _{\rm D} ,
\nonumber \\
\end{eqnarray}
where
\begin{equation}
\xi _{nk}=(-1)^k \sqrt{\binom{n}{k} } \left( \sqrt{T} \right)^{n-k}
\left( \sqrt{R} \right)^k ,
\end{equation}
and $\binom{n}{k} $ is the binomial coefficient and $R=1-T$ is the
reflectance.

\subsection{Photon number resolving detector case}

When $k$ photons are detected in the beam at path C and $l$ are
detected in path D by ideal photon number resolving detectors, the
conditional state is given by
\begin{eqnarray}
|\psi _{\rm NG}^{(kl)} \rangle _{\rm AB} &\propto &{_{\rm D} \langle
}l|{_{\rm C} \langle }k
|\psi \rangle _{\rm ABCD} \nonumber \\
&=&\sum _{n=\max \{k,l\} }^\infty \alpha _n \xi _{nk} \xi _{nl}
|n-k\rangle _{\rm A} |n-l\rangle _{\rm B} , \nonumber \\
\end{eqnarray}
where $|\psi _{\rm NG} \rangle _{\rm AB} $ is still a pure state. In
the case of $k=l=1$,
\begin{eqnarray}
|\psi _{\rm NG}^{(1)} \rangle _{\rm AB}
&=&\frac{1}{\sqrt{P_{\rm det}^{(1)} } }
\sum _{n=0}^\infty \alpha _{n+1} \xi _{(n+1),1}^2
|n\rangle _{\rm A} |n\rangle _{\rm B} \nonumber \\
&=&\sum _{n=0}^\infty c_n^{(1)} |n\rangle _{\rm A} |n\rangle _{\rm
B} , \label{output_NG}
\end{eqnarray}
where
\begin{eqnarray}
P_{\rm det}^{(1)} &=&\sum _{n=0}^\infty |\alpha _{n+1} |^2
|\xi _{(n+1),1}^2 |^2
\nonumber \\
&=&\frac{(1-\lambda^2 )\lambda^2 T^2 (1+\lambda^2 T^2 )} {(1-\lambda
^2 T^2 )^3 } \left( \frac{R}{T} \right)^2 \label{prob_pure_NG}
\end{eqnarray}
is the probability of detecting one photon in each arm. The state
(\ref{output_NG}) is \textit{not} Gaussian any more.

\subsection{On/off type detector case}

An ideal on/off type detector is described by a positive operator-valued
measure (POVM) with elements
\begin{equation}
\left\{
\begin{array}{lcl}
\hat{\Pi }^{(\rm off)} &=&|0\rangle \langle 0|, \\
\hat{\Pi }^{(\rm on)} &=&|1\rangle \langle 1|+|2\rangle \langle 2|
+\cdots =\hat{1}-|0\rangle \langle 0|.
\end{array}
\right.
\end{equation}
The two-mode squeezed state is transformed into a mixed non-Gaussian state
\begin{eqnarray}
\hat{\rho }_{\rm NG} &=&\frac{{\rm Tr}_{\rm CD} \left[ |\psi \rangle
_{\rm (ABCD)} \langle \psi | \otimes \left( \hat{\Pi }^{\rm (on)}
_{\rm C} \otimes \hat{\Pi }^{\rm (on)} _{\rm D} \right) \right] }
{\mathcal{P}_{\rm det} } \nonumber \\
&=&\frac{1}{\mathcal{P}_{\rm det} } \sum _{i,j=1}^\infty |\Phi _{ij}
\rangle _{\rm (AB)} \langle \Phi _{ij} |, \label{mixed_NG}
\end{eqnarray}
where
\begin{equation}
|\Phi _{ij} \rangle _{\rm AB} =\sum _{n=\max \{ i,j\} }
\alpha _n \xi _{ni} \xi _{nj} |n-i\rangle _{\rm A} |n-j\rangle _{\rm B} ,
\end{equation}
and $\mathcal{P}_{\rm det} $ is the probability of detecting at
least one photon in each of the paths C and D,
\begin{eqnarray}
\mathcal{P}_{\rm det} &=&{\rm Tr}_{\rm ABCD} \left[ |\psi \rangle
_{\rm (ABCD)} \langle \psi | \otimes \left( \hat{\Pi }^{\rm (on)}
_{\rm C}
\otimes \hat{\Pi }^{\rm (on)} _{\rm D} \right) \right] \nonumber \\
&=&\frac{\lambda^2 (1-T)^2 (1+\lambda^2 T)} {(1-\lambda^2
T)(1-\lambda^2 T^2 )} . \label{prob_mixed_NG}
\end{eqnarray}

The mean photon number of the state $\hat{\rho } _{\rm NG} $ is
expressed as
\begin{eqnarray}
\bar{N}_{\rm NG} &=&\textrm{Tr}_{\rm AB} \left[
\hat{\rho }_{\rm NG} \otimes \left( \hat{N}_{\rm A} +\hat{N} _{\rm B} \right)
\right] \nonumber \\
&=&\frac{2(1-\lambda^2 )}{\mathcal{P} _{\rm det} } \Bigg[
\frac{\lambda^2 T}{(1-\lambda^2 )^2 }
-\frac{\lambda^2 T}{(1-\lambda^2 T)^2 } \nonumber \\
&&\hspace{10mm} -\frac{\lambda^2 T^2 }{(1-\lambda^2 T)^2 }
+\frac{\lambda^2 T^2 }{(1-\lambda^2 T^2 )^2 } \Bigg] ,
\end{eqnarray}

\begin{figure}
\centering
\includegraphics[bb=55 95 560 740, angle=-90, width=\linewidth]{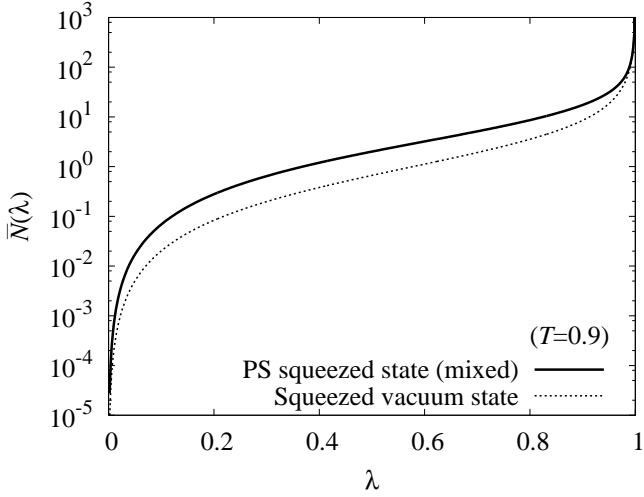}
\caption{\label{mean_n} Mean photon number
of the photon-subtracted mixed state
(thick solid line) and original two-mode squeezed vacuum state
(dotted line). }
\end{figure}

In Fig.~\ref{mean_n}, the mean photon number $\bar{N}_{\rm NG}$ is
shown as a function of $\lambda$, with that of the two-mode squeezed
vacuum state, $\bar{N}_{\rm SQ} =2\lambda^2 /(1-\lambda^2 ) $, for
comparison. The transmittances of the tapping beam splitters are
chosen as $T=0.9$.

The increase of the mean photon number for the photon-subtracted
squeezed state is due to the fact that the generation process is
based on the event selection of the components of higher numbers of
photons by excluding the original vacuum component of the input
two-mode squeezed state.

\section{Computable entanglement measures} \label{monotonicity}

In this section, we briefly review the negativity and the
logarithmic negativity as computable entanglement measures that
possess the properties of an entanglement monotone \cite{Vidal02}:
(i) The entanglement measure $E$ is a non-negative functional,
$E(\hat{\rho } )\geq 0$, (ii) if $\hat{\rho } $ is separable,
$E(\hat{\rho } )=0$,
and (iii) $E(\hat{\rho } )$ does not increase \textit{on average}
under the LOCC.

The negativity of a bipartite mixed state $\hat{\rho}$, denoted by
${\cal N}(\hat{\rho})$, is defined as the absolute value of the sum
of the negative eigenvalues of $\hat{\rho}^{PT}$, the partial
transpose of $\hat{\rho}$ with respect to either subsystem.  We may
write this as
\begin{equation}
\label{neg} {\cal
N}(\hat{\rho})=\frac{1}{2}\mathrm{Tr}\left(\sqrt{(\hat{\rho}^{PT})^{2}}-\hat{\rho}^{PT}\right)=\frac{||\hat{\rho}^{PT}||-1}{2},
\end{equation}
where $||{\cdot}||$ denotes the trace-norm.  It is quite easy to
prove that ${\cal N}{\geq}0$.  For a separable state,
$\hat{\rho}=\sum_{i}p_{i}\hat{\rho }_{{\rm A} i}\otimes \hat{\rho
}_{{\rm B} i}$, the partial transpose with respect to either
subsystem, say B, is given by $\hat{\rho}=\sum_{i}p_{i}\hat{\rho
}_{{\rm A} i}\otimes \hat{\rho}^{T}_{{\rm B} i}$.  This is also a
state, and it therefore has zero negativity. Furthermore, under
LOCC, $\mathcal{N}$ does not increase on average
\cite{Vidal02,Eisert_Ph.D_Thesis}.  It is also convex, i.e.
\begin{equation}
\sum _i p_i \mathcal{N} (\hat{\rho } _i )\leq \mathcal{N}(\hat{\rho
} ),
\end{equation}
where ${\hat\rho}=\sum_{i}p_{i}{\hat\rho}_{i}$.  The logarithmic
negativity is defined as
\begin{equation}
E_\mathcal{N}(\hat{\rho})=\log _2(1+2{\cal
N}(\hat{\rho}))={\log}_{2}(||\hat{\rho}^{PT}||). \label{logarithmic
negativity}
\end{equation}
In addition to the properties (i), (ii), and (iii), this quantity is
additive because $||\hat{\rho }^{PT} \otimes \hat{\sigma }^{PT} ||
=||\hat{\rho }^{PT} ||\cdot ||\hat{\sigma }^{PT} ||$, which is also
a desired property of a good entanglement measure.

For a pure entangled state
\begin{equation}
|\chi \rangle _{\rm AB} =\sum _n c_n |n\rangle _{\rm A}
|n\rangle _{\rm B},
\end{equation}
the negativity and the logarithmic negativity can be calculated
analytically, where, without loss of generality, we take the
Schmidt coefficients $c_n $ to be non-negative. We use the fact that
\begin{equation}
\left| \left| \left\{ |\chi \rangle _{\rm (AB)} \langle \chi
|\right\}^{PT} \right| \right| =\left( \sum _n c_n \right)^2 ,
\label{neg_formula}
\end{equation}
and we can obtain \cite{Vidal02}:
\begin{eqnarray}
{\cal N}(|\chi \rangle )&=&\frac{\left( \sum _n c_n \right) ^2 -1}{2},
\label{negschmidt} \\
E_\mathcal{N}(|\chi \rangle )&=&2{\log}_{2}\left( \sum _n c_n
\right) . \label{lognegschmidt}
\end{eqnarray}
For example, one can calculate those of the squeezed vacuum state (\ref{SVS})
as
\begin{eqnarray}
{\cal N}(|r^{(2)} \rangle )&=&\frac{{\lambda}}{1-{\lambda}}, \label{svneg} \\
E_\mathcal{N}(|r^{(2)} \rangle )&=&{\log}_{2}(1+{\lambda})-{\log}_{2}(1-{\lambda}),
\label{svlogneg}
\end{eqnarray}
where we have used the Schmidt coefficients described in Eq. (\ref{Schmidt}).

As another example, we here consider the ideal limit of photon-subtracted
squeezed state ($T\rightarrow 1$). In this limit, the photon-subtracted
squeezed state with on/off detector $\hat{\rho }_{\rm NG} $
is exactly identical to the pure state case $|\psi _{\rm NG} ^{(1)} \rangle
\langle \psi _{\rm NG} ^{(1)} |$, although the detection probability
$P_{\rm det} ^{(1)} $ approaches zero. From Eq. (\ref{output_NG}),
\begin{eqnarray}
\lim _{T\rightarrow 1} \sum _{n=0} ^\infty c_n ^{(1)}
&=&\sqrt{\frac{(1-\lambda ^2 )^3 }{1+\lambda ^2 } }
\sum _{n=0} ^\infty \lambda ^n (n+1) \nonumber \\
&=&\sqrt{\frac{(1-\lambda ^2 )^3 }{1+\lambda ^2 } }
\left( \lambda \frac{d}{d\lambda } \sum _{n=0} ^\infty \lambda ^n 
+\sum _{n=0} ^\infty \lambda ^n \right) \nonumber \\
&=&\sqrt{\frac{(1+\lambda )^3 }{(1+\lambda ^2 )(1-\lambda )} } ,
\end{eqnarray}
and thus we have
\begin{eqnarray}
\lim _{T\rightarrow 1} \mathcal{N} (|\psi _{\rm NG} ^{(1)} \rangle )
&=&\frac{\lambda (2+\lambda +\lambda ^2) }{(1+\lambda ^2 )(1-\lambda )} , \\
\lim _{T\rightarrow 1} E_\mathcal{N} (|\psi _{\rm NG} ^{(1)} \rangle )
&=&\log _2 \frac{(1+\lambda )^3 }{(1+\lambda ^2 )(1-\lambda )} .
\label{lim_log_neg}
\end{eqnarray}
As a consequence, we find that the logarithmic negativity
of the photon-subtracted squeezed state is always better than
that of the original squeezed vacuum for nonzero squeezing, i.e.
\begin{equation}
E_\mathcal{N} (|r^{(2)} \rangle )
\leq \lim _{T\rightarrow 1} E_\mathcal{N} (|\psi _{\rm NG} ^{(1)} \rangle ),
\label{relation_lim}
\end{equation}
where the equal sign is valid for $\lambda=0$.

In realistic situations, on the other hand, the photon subtraction
would be done by on/off detectors and the tapping beam splitters
with $T<1$. Then the generated states are inevitably reduced to
mixed states and it is typically not possible to obtain analytical
expressions for their negativity and logarithmic negativity.
However, as shown in the next section, one can often compute them
numerically using only linear algebra packages.

\section{Numerical evaluation of negativities}
\label{numerical result}

In this section, we explain the procedure for numerical evaluation
of the negativities of the photon-subtracted squeezed state with the on/off
detectors. First, we expand the output non-Gaussian state
(\ref{mixed_NG}) in the Fock basis,
\begin{equation}
\hat{\rho }_{\rm NG} =\sum _{m_1 ,m_2 ,n_1 ,n_2 } \rho _{m_1 m_2 n_1
n_2} |m_1 \rangle _{\rm A} \langle m_2 |
\otimes |n_1 \rangle _{\rm B} \langle n_2 |,
\end{equation}
where
\begin{eqnarray}
\lefteqn{ \rho _{m_1 m_2 n_1 n_2} } \nonumber \\
&=&{_{\rm B} \langle } n_1 | {_{\rm A} \langle } m_1 |\hat{\rho }_{\rm NG}
|m_2 \rangle _{\rm A} |n_2 \rangle _{\rm B} \nonumber \\
&=&\frac{1}{\mathcal{P}_{\rm det} } \sum _{i,j=1}^\infty
\alpha _{m_1 +i} \alpha _{m_2 +i} \xi _{(m_1 +i),i} \xi _{(m_1 +i),j} \nonumber \\
&&\times \xi _{(m_2 +i),i} \xi _{(m_2 +i),j} \delta _{(m_1 -n_1 ),(j-i)}
\delta _{(m_2 -n_2 ),(j-i)} , \nonumber \\
\end{eqnarray}
which means that the density matrix elements are zero unless
$m_1 -n_1 =m_2 -n_2 $.

\begin{figure}
\centering
\includegraphics[bb=55 55 555 470, width=0.9\linewidth]{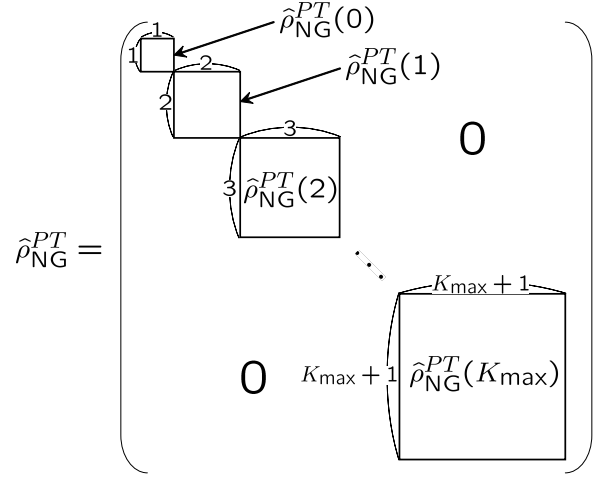}
\caption{\label{PT_matrix2} Partially-transposed density matrix
of the photon-subtracted mixed state
with the cut-off $K_{\rm max}$, which is block diagonal in Fock state basis. }
\end{figure}

The partial transpose of this state, with respect to mode B, is
\begin{equation}
\hat{\rho}_{\rm NG}^{PT}=\sum _{m_1 ,m_2 ,n_1 ,n_2 } \rho _{m_1 m_2 n_2
n_1} |m_1 \rangle _{\rm A} \langle m_2 | \otimes |n_1 \rangle _{\rm
B} \langle n_2 |,
\end{equation}
where the elements are zero unless $m_1 +n_1 =m_2 +n_2=K$. The
parameter $K$ is the total photon number of both beams at paths A and B, and
the partially-transposed density operator is block diagonal in Fock state basis,
where the blocks correspond to $K=0,1,2,{\ldots}$:
\begin{equation}
\hat{\rho }_{\rm NG}^{PT} =\bigoplus _{K=0}^\infty \hat{\rho }_{\rm
NG}^{PT}(K).
\end{equation}
Here, $\hat{\rho }_{\rm NG}^{PT}(K)$ is the $K$-th submatrix which
is a $(K+1)\times (K+1)$ real matrix (Fig. \ref{PT_matrix2}).

The eigenvalues are obtained by numerically diagonalizing
the partially transposed density matrix,
\begin{equation}
\hat{\varrho }=\hat{U}^T \hat{\rho }_{\rm NG}^{PT}\hat{U} ,
\end{equation}
which can separately be done for every submatrix one by one,
\begin{eqnarray}
\hat{\varrho}_K&=&\hat{U}_K^T\hat{\rho}_{\rm NG}^{PT}(K) \hat{U}_K.
\nonumber
\\
&=&\sum_{l=0}^K{\omega}_{l}^K|K-l{\rangle}_{\rm A} \langle K-l|
\otimes |l\rangle _{\rm B} \langle l|,
\end{eqnarray}
where $\omega _l^K $ is the $l$-th eigenvalue of $K$-th submatrix,
and
\begin{equation}
\hat{U}=\bigoplus_{K=0}^{\infty} \hat{U}_K .
\end{equation}
In the numerical calculation, we introduce a cut-off $K_{\rm max}$,
which must be large enough compared with the mean photon number.
Then, we sum up all the negative eigenvalues, which gives the
negativity,
\begin{equation}
\mathcal{N}(\hat{\rho }_{\rm NG} )=\frac{1}{\Delta _{K_{\rm max} } }
\sum _{K=0}^{K_{\rm max} } \sum _{\omega _l ^K <0} |\omega _l^K |,
\end{equation}
where
\begin{equation}
\Delta _{K_{\rm max} } (\lambda ) = \sum _{K=0}^{K_{\rm max} }
\textrm{Tr} \hat{\rho }_{\rm NG}^{PT}(K) \label{trace}
\end{equation}
is the trace of the truncated density matrix.
With this result, we can obtain the logarithmic negativity
straightforwardly using Eq. \eqref{logarithmic negativity}.

Throughout this paper, we set $T=0.9$ unless otherwise indicated.
In our computations, we chose a cut-off of $K_{\rm max}=50$. The
trace (\ref{trace}) could be used as a measure of the validity of
our chosen cut-off. For $\lambda =0.78$ and 0.88, which corresponds
to $\bar{N}_{\rm NG} \simeq 7.71$ and 14.7, $\Delta _{K_{\rm max}
=50} \simeq 1.000$ and 0.995, respectively. For the latter case,
indeed, it still misses $0.5\%$ of the support, but this precision
suffices for our analysis. In this connection, with $K_{\rm max}
=50$, it takes 3.5 days to calculate the negativity for a certain
$\lambda $, by means of a Mathematica program on Pentium 4 3.2E GHz
PC (including the generation of matrix components). It is necessary
to reach a satisfactory compromise between the precision of the
numerical calculation and the calculation cost. The numerical
calculation becomes progressively time-consuming with increasing
$K_{\rm max}$.

\begin{figure}
\centering
\includegraphics[bb=60 60 560 730, angle=-90, width=\linewidth]{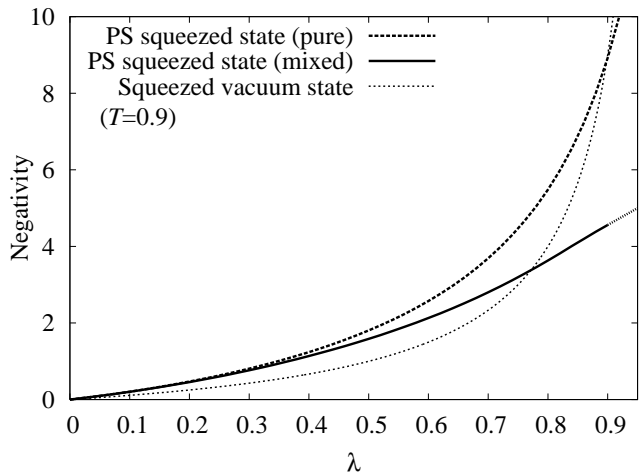}
\caption{\label{Negativity3}
Negativity of the photon-subtracted squeezed states,
in pure state case (thick dashed line) and mixed state case (thick solid line).
The dotted line corresponds to the input squeezed vacuum. }
\end{figure}

\begin{figure}
\centering
\includegraphics[bb=60 60 560 730, angle=-90, width=\linewidth]{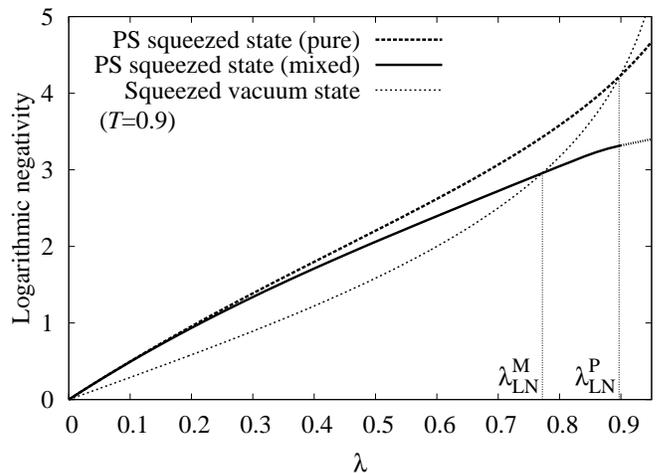}
\caption{\label{Log_Negativity3}
The same plot as Fig. \ref{Negativity3} of
the logarithmic negativity. The values of each intersection are
$\lambda _{\rm LN} ^{\rm P} \simeq 0.897$
and $\lambda _{\rm LN} ^{\rm M} \simeq 0.772$. }
\end{figure}

Figures \ref{Negativity3} and \ref{Log_Negativity3} show the
numerical values of the negativity and the logarithmic negativity,
respectively. The thick dashed line represents the photon-subtracted
squeezed state of the pure state case (\ref{output_NG}), while the
thick solid line corresponds to the mixed one (\ref{mixed_NG}). The
dotted line is for the input two-mode squeezed vacuum state
(\ref{svlogneg}). As can be seen in the figures, the photon
subtracted non-Gaussian states have a larger amount of entanglement
than the input two-mode squeezed vacuum state in a practical
squeezing range, $\lambda \lesssim \lambda _{\rm LN} ^{\rm P}
=0.897$ for the pure state case, and $\lambda \lesssim \lambda _{\rm
LN} ^{\rm M} =0.772$ for the mixed one, which correspond to 8.9 dB
and 7.1 dB ideal squeezing. Exceeding these $\lambda _{\rm LN} $'s,
however, the merit of non-Gaussian operation disappears. This is in
contrast to the case of $T\rightarrow 1$, where the curve for the
photon-subtracted squeezed state is always upper than that of the
squeezed vacuum state, as mentioned in Eq. (\ref{relation_lim}). For
$\lambda \lesssim 0.2$, that is, $\bar{N}_{\rm NG} \lesssim 0.28$,
the difference between the cases of the photon number resolving
detector and the on/off type detector is almost negligible compared
with the difference between them in the case of ideal two-mode
squeezed state. This means that the photon subtraction by a beam
splitter of transmittance $T=0.9$ and the on/off detectors emulate
the single photon subtraction well.

\section{Operational entanglement measures} \label{op_meas}

In this section, we review the previous results on the operational entanglement measures,
and compare them with the above results of logarithmic negativity.

\subsection{CV teleportation fidelity}

\begin{figure}
\centering
\includegraphics[bb=50 55 580 450, width=.9\linewidth]{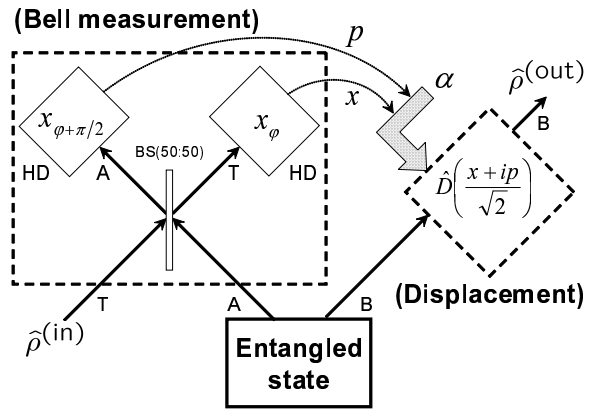}
\caption{\label{CV_teleportation} CV teleportation scheme. BS and HD
mean the beam splitter and homodyne detection, respectively. }
\end{figure}

The first operational measure is the CV teleportation fidelity \cite{Olivares03}.
In quantum teleportation, an unknown quantum state can be transferred
using entanglement and a classical channel.
The schematic is shown in Fig. \ref{CV_teleportation}.
First, the sender Alice and the receiver Bob share the entangled state
$\hat{\rho }_{\rm AB} ^{\rm (E)} $. Alice then performs the Bell measurement,
that is, a projective measurement in the maximally entangled basis
\begin{equation}
|\Pi (x,p)\rangle _{kl} =\frac{1}{\sqrt{2\pi } }
\int _{-\infty } ^\infty dy e^{ipy} |x+y\rangle _k |y\rangle _l \label{Bell}
\end{equation}
upon the unknown state $\hat{\rho }_{\rm T} ^{\rm (in)} $
and her fragment of entangled state (A). She obtains a couple
of measurement results $(x,p)$, and Bob's part of the initial state (B)
is correspondingly transformed into
\begin{equation}
\hat{\sigma } _{\rm B} (x,p)=\frac{ {_{\rm TA} \langle }\Pi (x,p)|
\left[ \hat{\rho }_{\rm T} ^{\rm (in)} \otimes \hat{\rho }_{\rm AB} ^{\rm (E)} \right]
|\Pi (x,p) \rangle _{\rm TA} }{P^{\rm (tlp)} (x,p)} ,
\end{equation}
where
\begin{equation}
P^{\rm (tlp)} (x,p)=\textrm{Tr} _{\rm B}
\left[ {_{\rm TA} \langle }\Pi (x,p)|
\left[ \hat{\rho }_{\rm T} ^{\rm (in)}
\otimes \hat{\rho }_{\rm AB} ^{\rm (E)} \right]
|\Pi (x,p) \rangle _{\rm TA} \right]
\end{equation}
is the Bell measurement probability distribution.
Finally, Bob corrects his state with the unitary transformation, that is,
the displacement operation
$\hat{D}(\alpha )=\exp [\alpha \hat{a}^\dagger -\alpha ^\ast \hat{a} ]$
according to Alice's measurement result,
\begin{equation}
\hat{\rho }_{\rm B} ^{\rm (out)} (x,p)
=\hat{D} _{\rm B} \left( \frac{x+ip}{\sqrt{2}}  \right) \hat{\sigma } _{\rm B} (x,p)
\hat{D} _{\rm B} ^\dagger \left( \frac{x+ip}{\sqrt{2}}  \right) .
\end{equation}
The average fidelity between the input and output is
\begin{equation}
\bar{F}=\int _{-\infty } ^\infty dx \int _{-\infty } ^\infty dp
P^{\rm (tlp)} (x,p)\textrm{Tr} \left[ \hat{\rho } ^{\rm (in)}
\hat{\rho }^{\rm (out)} (x,p)\right] ,
\end{equation}
which depends upon the input state $\hat{\rho }_{\rm T} ^{\rm (in)} $.
Let us consider that the input state is the coherent state,
\begin{equation}
\hat{\rho }_{\rm T} ^{\rm (in)} =|\alpha _0 \rangle _{\rm T} \langle \alpha _0 |,
\end{equation}
which is one of the unaffected state.
With the two-mode squeezed vacuum state (\ref{SVS}), the average fidelity is
\begin{equation}
\bar{F}_{\rm SQ} (\lambda )=\frac{1+\lambda }{2} ,
\end{equation}
and, with the non-Gaussian pure state (\ref{output_NG})
and mixed one (\ref{mixed_NG}),
\begin{eqnarray}
\bar{F}_{\rm NG} ^{(1)} (\lambda )&=&
\frac{(1-\lambda ^2 )\lambda ^2 T^2
\left( 1-\lambda T+\frac{\lambda ^2 T^2 }{2} \right) }
{2P_{\rm det} ^{(1)} (1-\lambda T)^3 } \left( \frac{R}{T} \right) ^2 , \nonumber \\
&& \label{fidelity_pure_NG} \\
\bar{\mathcal{F} }_{\rm NG} (\lambda )&=&\mathcal{F} _{11} (\lambda )
-\mathcal{F} _{10} (\lambda )-\mathcal{F} _{01} (\lambda )
+\mathcal{F} _{00} (\lambda ) \label{fidelity_mixed_NG},
\end{eqnarray}
respectively, where
\begin{equation}
\mathcal{F} _{ij} (\lambda )=
\frac{1}{2\mathcal{P}_{\rm det} } \frac{1-\lambda ^2 }
{1-\lambda T-\lambda ^2 \gamma _i \gamma _j
-\frac{\lambda ^2 T}{2} (\gamma _i +\gamma _j )}
\end{equation}
and $\gamma _1 =R$, $\gamma _0 =0$. See appendices \ref{appendix_A}
and \ref{appendix_B} for more detailed derivations.

\begin{figure}
\centering
\includegraphics[bb=65 60 555 735, angle=-90, width=.9\linewidth]{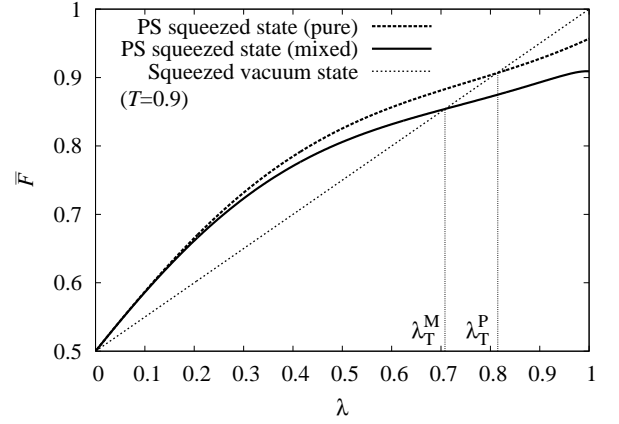}
\caption{\label{fidelity} Average fidelity of CV teleportation of coherent state,
with the photon-subtracted pure state (thick dashed line),
mixed one (thick solid line), and original
squeezed vacuum state (dotted line). The values of each intersection are
$\lambda _{\rm T} ^{\rm P} \simeq 0.815$
and $\lambda _{\rm T} ^{\rm M} \simeq 0.708$. }
\end{figure}

In the limit as $T\rightarrow 1$, $\bar{\mathcal{F}} _{\rm NG}
(\lambda ) =\bar{F}_{\rm NG} ^{(1)} (\lambda ) $ and
\begin{equation}
\bar{F}_{\rm SQ} (\lambda )\leq \lim _{T\rightarrow 1} \bar{F}_{\rm NG} ^{(1)} (\lambda ) ,
\end{equation}
where the equal sign is valid when $\lambda =0$ and 1. For $T<1$,
however, the average fidelities (\ref{fidelity_pure_NG}) and
(\ref{fidelity_mixed_NG}) do not attain unity even as $\lambda
\rightarrow 1$.  Figure \ref{fidelity} shows the average fidelities
for $T=0.9$.
We can see that the photon-subtracted squeezed states, in both the
pure and mixed state cases, are superior to the original squeezed
vacuum state for $\lambda \lesssim \lambda _{\rm T} ^{\rm P} =0.815$
and $\lambda \lesssim \lambda _{\rm T} ^{\rm M} =0.708$,
respectively. In the range of $\lambda _{\rm T} ^{\rm P} <\lambda $,
on the other hand, the squeezed vacuum state shows the best
performance, which is quantitatively consistent with the logarithmic
negativity result. As seen by comparing the results of Fig.
\ref{Log_Negativity3} and \ref{fidelity}, whenever the fidelity is
improved by the photon subtraction, so is the logarithmic
negativity, i.e. $\lambda _{\rm T} <\lambda _{\rm LN} $. We cannot,
however, exclude the possibility that this is not true for the other
input states to be teleported. To clarify the exact relation between
the improvements of the logarithmic negativity and the teleportation
fidelity, one has to optimize every component of teleportation
protocol, such as measurements, etc., for the photon-subtracted
entangled resource over all possible input states. These might be a
highly non-trivial task.

\subsection{Mutual information in CV dense coding scheme}

\begin{figure}
\centering
\includegraphics[bb=55 55 570 405, width=.9\linewidth]{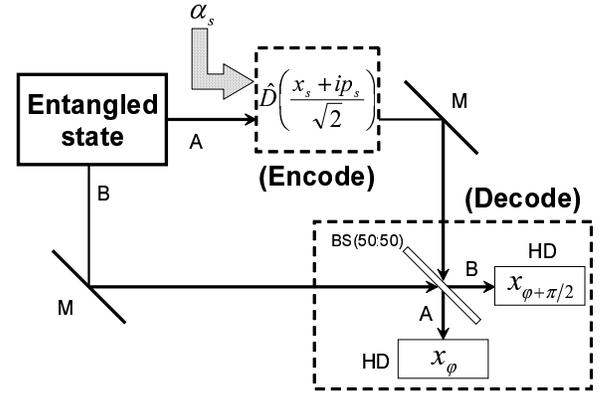}
\caption{\label{CV_coding} CV dense coding scheme scheme, with power of
signal modulation $\alpha _s $. BS, M, and HD
mean beam splitter, mirror, and homodyne detection, respectively. }
\end{figure}

The second operational measure relates to the CV dense coding scheme
\cite{Kitagawa05}. The schematic is illustrated in Fig.
\ref{CV_coding}.
In this scheme, Alice and Bob share the entangled state $\hat{\rho
}_{\rm AB} ^{\rm (E)} $ initially. Then, Alice encodes a
\textit{classical} message by a modulation on the beam path A,
\begin{equation}
\hat{U}_{\rm A} (x_s ,p_s )
=e^{-(i/2)x_s p_s } \hat{D} \left( \frac{x_s +ip_s }{\sqrt{2} } \right)
=e^{-ip_s \hat{x} _{\rm A} } e^{ix_s \hat{p}_{\rm A} } ,
\end{equation}
where $x_s $ and $p_s $ are related to the power of signal
modulation $\alpha _s $.  For the sake of simplicity, we consider
the quaternary phase-shift keying (QPSK), with equal likelihood:
$a_{00} =(x_s =\sqrt{2}\beta ,p_s =\sqrt{2}\beta )$, $a_{01}
=(\sqrt{2}\beta ,-\sqrt{2}\beta )$, $a_{10} =(-\sqrt{2}\beta
,\sqrt{2}\beta )$, $a_{11} =(-\sqrt{2}\beta ,-\sqrt{2}\beta )$ for
some real $\beta $, and $P(a_{kl} )=1/4$. Bob decodes the signals by
the Bell measurement (\ref{Bell}), and the decision rule is as
follows: $\{b_{00} =(x\geq 0,p\geq 0),b_{01} =(x\geq 0,p<0), b_{10}
=(x<0,p\geq 0), b_{11} =(x<0,p<0)\}$. The channel matrix of CV dense
coding $P^{\rm (ch)} (b_{mn} |a_{kl} )$, whose elements are the
conditional probabilities, is calculated with the homodyne
probability distribution,
\begin{eqnarray}
\lefteqn{P^{\rm (HD)} (x,p|x_s ,p_s ) } \nonumber \\
&=&{_{\rm AB} \langle } \Pi (x,p)|\hat{U}_{\rm A} (x_s ,p_s )
\hat{\rho }_{\rm AB} ^{\rm (E)}
\hat{U} _{\rm A} ^\dagger (x_s ,p_s )|\Pi (x,p)\rangle _{\rm AB} . \nonumber \\
&& \label{Ch-Mtrx}
\end{eqnarray}
For example, $P^{\rm (ch)} (b_{00} |a_{00})$ element of the channel
matrix is
\begin{equation}
P^{\rm (ch)} (b_{00} |a_{00} )=\int _0 ^\infty dx \int _0 ^\infty dp
P^{\rm (HD)} (x,p|\sqrt{2} \beta ,\sqrt{2} \beta ),
\end{equation}
and the other components are calculated similarly. With this
$4\times 4$ channel matrix, the mutual information is calculated as
\begin{eqnarray}
I(\textrm{A;B})&=&\sum _{k,l,m,n} P(a_{kl} )P^{\rm (ch)} (b_{mn} |a_{kl} ) \nonumber \\
&&\times \log _2 \left[ \frac{P^{\rm (ch)} (b_{mn} |a_{kl} )}{\sum _{k' ,l' }
P(a_{k' l' } )P^{\rm (ch)} (b_{mn} |a_{k' l' } )} \right] \ \textrm{[bit]}. \nonumber \\
\end{eqnarray}
With the two-mode squeezed vacuum state (\ref{SVS}),
\begin{eqnarray}
\lefteqn{I_{\rm SQ} (\textrm{A;B})} \nonumber \\
&&=\left[ 1+\textrm{erf}\left( \sqrt{\frac{1+\lambda }
{1-\lambda } } \beta \right) \right]
\log _2 \left[ 1+\textrm{erf}\left( \sqrt{\frac{1+\lambda }
{1-\lambda } } \beta \right) \right] \nonumber \\
&&+\left[ 1-\textrm{erf}\left( \sqrt{\frac{1+\lambda }
{1-\lambda } } \beta \right) \right]
\log _2 \left[ 1-\textrm{erf}\left( \sqrt{\frac{1+\lambda }
{1-\lambda } } \beta \right) \right] , \nonumber \\
\end{eqnarray}
where
\begin{equation}
\textrm{erf}(x)=\frac{2}{\sqrt{\pi } } \int _0 ^x dt e^{-t^2 }
\end{equation}
is the error function. With the non-Gaussian pure state
(\ref{output_NG}),
\begin{eqnarray}
I_{\rm NG} ^{(1)} (\textrm{A;B})&=&\frac{1}{4} (I_1 \log _2 I_1
+I_2 \log _2 I_2 \nonumber \\
&&\hspace{10mm} +I_3 \log _2 I_3
+I_4 \log _2 I_4 ) , \label{mutual_info_pure_NG}
\end{eqnarray}
and
\begin{eqnarray}
I_1 &=&\mathcal{D}_\mu \frac{1}{\mu }
\left[ 1+\textrm{erf}\left( \sqrt{\displaystyle \frac{1+\lambda T}{1-\lambda T} \mu }
\beta \right) \right] ^2 \Bigg| _{\mu =1} , \\
I_2 =I_3 &=&\mathcal{D}_\mu \frac{1}{\mu }
\left[ 1-\textrm{erf}\left( \sqrt{\displaystyle \frac{1+\lambda T}{1-\lambda T} \mu }
\beta \right) ^2 \right] \Bigg| _{\mu =1} , \\
I_4 &=&\mathcal{D}_\mu \frac{1}{\mu }
\left[ 1-\textrm{erf}\left( \sqrt{\displaystyle \frac{1+\lambda T}{1-\lambda T} \mu }
\beta \right) \right] ^2 \Bigg| _{\mu =1} ,
\end{eqnarray}
where
\begin{eqnarray}
\mathcal{D}_\mu &=&\frac{(1-\lambda ^2 )\lambda ^2 T^2 }{P_{\rm det} ^{(1)}
(1-\lambda T)^2 (1-\lambda ^2 T^2 )} \left(\frac{R}{T} \right) ^2 \nonumber \\
&&\times \left[ \left(\frac{\lambda T}{1+\lambda T} \right) ^2
\frac{\partial ^2 }{\partial \mu ^2 }
+\frac{2\lambda T}{1+\lambda T} \frac{\partial }{\partial \mu } +1\right] ,
\end{eqnarray}
containing the differential operation with respect to the auxiliary parameter $\mu $.
With the non-Gaussian mixed state (\ref{mixed_NG}),
\begin{eqnarray}
\mathcal{I} _{\rm NG} (\textrm{A;B})&=&\frac{1}{4} ( \mathcal{I}_1 \log _2 \mathcal{I}_1
+\mathcal{I}_2 \log _2 \mathcal{I}_2 \nonumber \\
&&\hspace{10mm} +\mathcal{I}_3 \log _2 \mathcal{I}_3
+\mathcal{I}_4 \log _2 \mathcal{I}_4 ) , \label{mutual_info_mixed_NG}
\end{eqnarray}
where
\begin{eqnarray}
\mathcal{I}_1 &=&\sum _{i,j=0} ^1 (-1)^{i+j} \mathcal{C} _{ij}
\left[ 1+\textrm{erf}(\Omega _{ij} \beta )\right] ^2 , \\
\mathcal{I}_2 =\mathcal{I}_3 &=&\sum _{i,j=0} ^1 (-1)^{i+j} \mathcal{C} _{ij}
\left[ 1-\textrm{erf}(\Omega _{ij} \beta )^2 \right] , \\
\mathcal{I}_4 &=&\sum _{i,j=0} ^1 (-1)^{i+j} \mathcal{C} _{ij}
\left[ 1-\textrm{erf}(\Omega _{ij} \beta )\right] ^2 ,
\end{eqnarray}
and
\begin{eqnarray}
\mathcal{C} _{ij} &=&\frac{1}{4\mathcal{P}_{\rm det} }
\frac{1-\lambda ^2 }{1-\lambda ^2 (T+\gamma _i )(T+\gamma _j )} , \\
\Omega _{ij} &=&\sqrt{\frac{1-\lambda ^2 (T+\gamma _i )(T+\gamma _j )}
{(1-\lambda T)^2 -\lambda ^2 \gamma _i \gamma _j } } ,
\end{eqnarray}
where $\gamma _1 =R$ and $\gamma _0 =0$ as before. See appendices
\ref{appendix_A} and \ref{appendix_B} for more detailed derivations.

\begin{figure}[t]
\centering
\includegraphics[bb=60 55 555 740, angle=-90, width=.9\linewidth]
{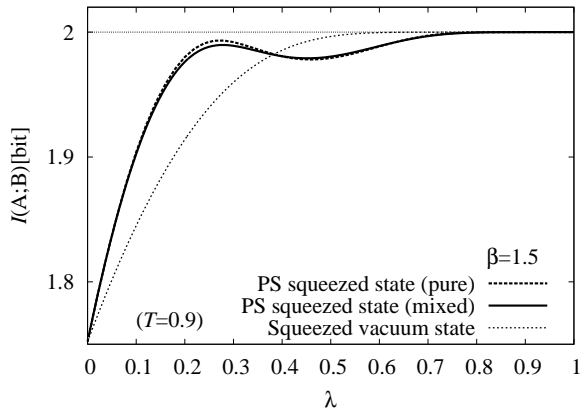}
\caption{\label{mutual_info_15} Mutual information in dense coding scheme
with the photon-subtracted pure state (dashed line),
mixed one (solid line),
and squeezed vacuum state (dotted line) at $\beta =1.5$. }
\end{figure}

\begin{figure}[t]
\centering
\includegraphics[bb=60 55 555 740, angle=-90, width=.9\linewidth]
{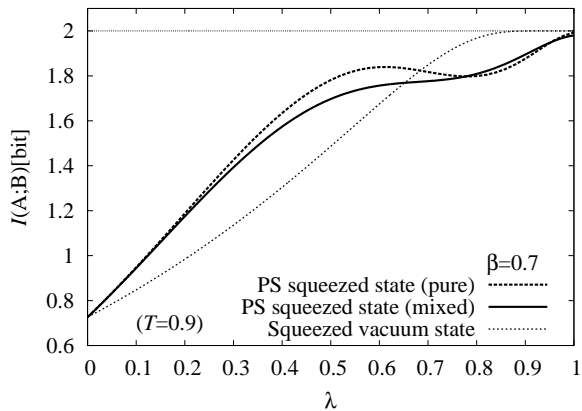}
\caption{\label{mutual_info_07} The same plot as Fig. \ref{mutual_info_15}
at $\beta =0.7$. }
\end{figure}

\begin{figure}[t]
\centering
\includegraphics[bb=60 50 555 740, angle=-90, width=.9\linewidth]
{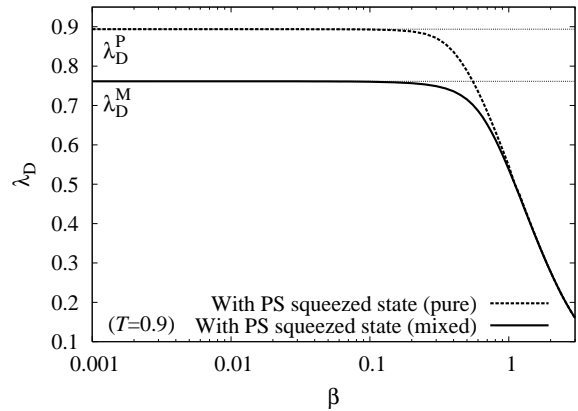}
\caption{\label{gain} Intersections of
the input squeezed state curve and photon-subtracted
squeezed state curves.
It converges to $\lambda _{\rm D} ^{\rm P} \simeq 0.894$
for the pure state case, and $\lambda _{\rm D} ^{\rm M} \simeq 0.762$
for mixed one, as $\beta \rightarrow 0$. }
\end{figure}

In the limit as $T\rightarrow 1$, $\mathcal{I}_{\rm NG}
(\textrm{A;B}) =I_{\rm NG} ^{(1)} (\textrm{A;B})$ which has no
intersection with $I_{\rm SQ} (\textrm{A;B})$ unless $\lambda =0$,
therefore
\begin{equation}
I_{\rm SQ} (\textrm{A;B})\leq \lim _{T\rightarrow 1}
I_{\rm NG} ^{(1)} (\textrm{A;B}).
\end{equation}
For $T<1$, however, above inequality cannot hold for larger $\lambda
$. In Figs. \ref{mutual_info_15} and \ref{mutual_info_07}, the
mutual information with the photon-subtracted squeezed states
($T=0.9$) and the squeezed vacuum state are shown, at $\beta =1.5$
and 0.7, respectively.
From these results, we can see that the range of $\lambda $ for
which the non-Gaussian operation brings the gain to the mutual
information becomes wide as $\beta $ becomes small. When $\beta $
gets smaller, the overlap between the probability distributions of
different signals increases, and the distinction of encoded signals
becomes more difficult. In this situation, we can see the
\textit{bona fide} effect of entanglement in CV dense coding scheme.
As $\beta \rightarrow 0$, the intersections of the curves for the
non-Gaussian states and the squeezed vacuum state approach $\lambda
_{\rm D} ^{\rm P} \simeq 0.894$ and $\lambda _{\rm D} ^{\rm M}
\simeq 0.762$, respectively (Fig. \ref{gain}).
These intersections are quite close to those of the logarithmic
negativity. This is consistent with the logarithmic negativity
result in the sense that the mutual information indicates the range
for which non-Gaussian operation enhances the entanglement.

\section{Discussion and conclusion} \label{discussion}

In this paper, we have studied how to quantify the entanglement of
the non-Gaussian mixed state which is generated by the photon
subtraction from the two-mode squeezed vacuum state. In order to
enhance the entanglement of a Gaussian state, shared by
spatially-separated parties, using LOCC, then non-Gaussian
operations are necessary. When the photon subtraction is carried out
using on/off detectors, the resulting state is generally a mixed
state. Evaluating the entanglement of such a state is far from
trivial.

We have applied the negativity and the logarithmic negativity, which
are known to be entanglement monotones, to this problem. We have
compared the gain in terms of these measures with the one in terms
of certain protocol-specific operational measures, namely the
teleportation fidelity and the mutual information of the dense
coding.

In the asymptotic limit $T\rightarrow 1$, one would have an ideal
single photon subtraction even with on/off detector and hence have
an pure non-Gaussian state, although the successful probability of
event selection approaches zero. This ideal non-Gaussian state is
always superior to the input squeezed vacuum state for all the
measures. For $T<1$, on the other hand, up to a certain point of
$\lambda $, the photon-subtracted squeezed states, both the pure and
mixed states, are superior to the input squeezed state with respect
to all the measures. Exceeding this point $\lambda $, the
non-Gaussian operation brings no gain, and the entangled squeezed
state shows the better performance. When $\lambda $ approaches
unity, the effect of the non-Gaussian operation for entanglement
enhancement gets lost. This is because the initial squeezed state
approaches the maximally entangled EPR state, as
${\lambda}{\rightarrow}1$, whose entanglement cannot be enhanced by
any physical process.

While the operational measures have clear physical meanings, they
directly depend on the input state characteristics, and the
evaluation based on them vary for the protocols. The (logarithmic)
negativity is, on the other hand, independent of any such external
parameters. This quantity reflects the entanglement as an intrinsic
property of the state of interest, namely the final output state.
We have found that whenever the enhancement is seen in terms of the
operational measures, the negativity and the logarithmic negativity
are also enhanced. For the dense coding scheme, in particular, the
upper limit of $\lambda $ below which the non-Gaussian operational
gain can be seen, $\lambda _{\rm D} ^{\rm P} $ (pure state case) and
$\lambda _{\rm D} ^{\rm M} $ (mixed state case) approach the ones
measured by the logarithmic negativity $\lambda _{\rm LN} ^{\rm P} $
and $\lambda _{\rm LN} ^{\rm M} $, respectively, as the modulation
signal power $\beta $ gets smaller. It would be interesting to
investigate whether the intersections $\lambda _{\rm D}$ in the
dense coding as well as $\lambda _{\rm T}$ for the teleportation are
identical to $\lambda _{\rm LN}$ or not after the operational
measures are further optimized with respect to all the possibilities
for external parameters of input states. In other words, the
logarithmic negativity, not containing any additional parameter,
might give the universal upper limit for the region of $\lambda $,
where the gain by the non-Gaussian operation can be seen.

The operational meaning of the logarithmic negativity
has not completely been clear yet.
The relationship between the logarithmic negativity and other
entanglement measures, such as the entanglement of formation, has
not yet been fully elucidated.  Restricted to the case of symmetric
Gaussian entangled states only, they always indicate the same
ordering \cite{Adesso05}. Generally, however, they could give the
different ordering for a few cases \cite{Adesso05,Eisert99}.
Our findings will provide some insight into
further studies of the evaluation of the performance of the
non-Gaussian operations.

For practical application of our results to
the laboratory experiment,
various imperfections should be considered, such as
the limited quantum efficiency and
nonzero dark count rate of the photodetector,
and linear loss in the optical paths.
Then the logarithmic negativity evaluation
in consideration of these imperfections is the important issue.
However, those effects cause
the complex mixing between two modes A and B.
Therefore, it is not obvious whether the partially-transposed
density matrix $\hat{\rho}^{PT} _{\rm NG} $ can be split into series of submatrices,
as in the case of the ideal setup. This would
make the analysis much harder even by numerical simulation.
Such analysis for more practical situations is a future problem.

\begin{acknowledgments}
The authors would like to M.~Ban for valuable discussions.
\end{acknowledgments}

\appendix

\section{Derivation of (\ref{fidelity_pure_NG}) and (\ref{mutual_info_pure_NG})}
\label{appendix_A}

Let us describe the non-Gaussian pure state in terms of coherent
states, which is suitable for inner products with both the Fock
state basis and continuous variable basis.
\begin{eqnarray}
\langle n|\alpha \rangle &=&e^{-\frac{|\alpha |^2 }{2} }
\frac{\alpha ^n }{\sqrt{n!} } , \\
\langle x|\alpha \rangle
&=&\frac{1}{\sqrt[4]{\pi } }
\exp \left[ -\frac{x^2 }{2} +\sqrt{2}x \alpha -\frac{\alpha ^2 }{2}
-\frac{|\alpha |^2 }{2} \right] .
\end{eqnarray}

First, we consider the average fidelity of CV teleportation.
With the coherent state basis, the non-Gaussian pure state (\ref{output_NG})
is expressed with an auxiliary parameter $\mu $,
\begin{eqnarray}
\lefteqn{ |\psi _{\rm NG} ^{(1)} \rangle _{\rm AB} } \nonumber \\
&=&\frac{\sqrt{1-\lambda ^2 } }{\pi ^2 \sqrt{P_{\rm det} ^{(1)} } }
\int d^2 \alpha \int d^2 \beta R\alpha \beta \nonumber \\
&&\times \exp \left[ -\frac{1+R}{2} (|\alpha |^2 +|\beta |^2 )
+\lambda \alpha ^\ast \beta ^\ast \right] \nonumber \\
&&\hspace{30mm} |\sqrt{T} \alpha \rangle _{\rm A} |\sqrt{T} \beta \rangle _{\rm B}
\nonumber \\
&=&\frac{R\sqrt{1-\lambda ^2 } }{\pi ^2 \sqrt{P_{\rm det} ^{(1)} } }
\frac{\partial }{\partial \mu } \int d^2 \alpha \int d^2 \beta \nonumber \\
&&\times \exp \left[ -\frac{1+R}{2} (|\alpha |^2 +|\beta |^2 )
+\lambda \alpha ^\ast \beta ^\ast +\mu \alpha \beta \right] \nonumber \\
&&\hspace{30mm} |\sqrt{T} \alpha \rangle _{\rm A} |\sqrt{T} \beta \rangle _{\rm B}
\Bigg| _{\mu =0} , \label{pure_NG_coherent}
\end{eqnarray}
where $P_{\rm det} ^{(1)} $ is given by (\ref{prob_pure_NG}), and
the auxiliary parameter $\mu $ should be set to zero
after all integration and differential operations.
The inner product between the input state $|\alpha _0 \rangle $
and unnormalized output state after teleportation operation is
\begin{eqnarray}
\lefteqn{{_{\rm B} \langle } \alpha _0 |
\sqrt{P_{\rm NG} ^{\rm (tlp)} } |\psi _{\rm out} \rangle _{\rm B} }\nonumber \\
&=&{_{\rm B} \langle } \alpha _0 | \sqrt{P_{\rm NG} ^{\rm (tlp)} }
\hat{D}_{\rm B} (\xi ) \left[ {_{\rm TA} \langle } \Pi (x,p)
|\alpha _0 \rangle _{\rm T} |\psi _{\rm NG} ^{(1)} \rangle _{\rm AB} \right]
\nonumber \\
&=&\frac{R\sqrt{1-\lambda ^2 } }{\sqrt{2\pi P_{\rm det} ^{(1)} } }
\frac{\partial }{\partial \mu } \frac{1}{1-\lambda \mu } \nonumber \\
&&\times \exp \left[ -\left( 1-\frac{\lambda T}{1-\lambda \mu } \right)
\left| Q \right| ^2 +\frac{i}{2} xp \right] \Bigg| _{\mu =0} ,
\end{eqnarray}
where
\begin{equation}
Q=\frac{x+ip}{\sqrt{2} } -\alpha _0 = \xi -\alpha _0. \label{Q}
\end{equation}
Therefore the $(x,p)$ component of the fidelity is
\begin{eqnarray}
\lefteqn{P_{\rm NG} ^{\rm (tlp)} F_{\rm NG} ^{(1)} (x,p) } \nonumber \\
&=&\frac{R^2 (1-\lambda ^2 )}{2\pi P_{\rm det} ^{(1)} }
\frac{\partial ^2 }{\partial \mu _1 \partial \mu _2 }
\frac{1}{(1-\lambda \mu _1 )(1-\lambda \mu _2 )} \nonumber \\
&&\times \exp \left[ -\left( 2-\frac{\lambda T}{1-\lambda \mu _1 }
-\frac{\lambda T}{1-\lambda \mu _2 } \right)
\left| Q\right| ^2 \right] \Bigg| _{\mu _1 =\mu _2 =0} . \nonumber \\
&& \label{fidelity_xp_pure_NG}
\end{eqnarray}
By integrating Eq. (\ref{fidelity_xp_pure_NG})
with respect to $x$ and $p$, we obtain the average fidelity,
\begin{eqnarray}
\lefteqn{\bar{F}_{\rm NG} ^{(1)} (\lambda )} \nonumber \\
&=&\int _{-\infty } ^\infty dx \int _{-\infty } ^\infty dp
P_{\rm NG} ^{\rm (tlp)} F_{\rm NG} ^{(1)} (x,p) \nonumber \\
&=&\frac{(1-\lambda ^2 )\lambda ^2 T^2
\left( 1-\lambda T+\displaystyle \frac{\lambda ^2 T^2 }{2} \right) }
{2P_{\rm det} ^{(1)}
(1-\lambda T)^3 } \left( \frac{R}{T} \right) ^2 ,
\end{eqnarray}
which is independent of the parameter $\alpha _0 $.

Then, we consider the mutual information of CV dense coding channel.
\begin{eqnarray}
\lefteqn{{_{\rm AB} \langle } \Pi (x,p)|\hat{U} _{\rm A} (x_s ,p_s )
|\psi _{\rm NG} ^{(1)} \rangle _{\rm AB} e^{-ip_s (x-x_s )} } \nonumber \\
&=&\sqrt{\frac{1-\lambda ^2 }{2\pi ^5 P_{\rm det} ^{(1)} } }
\int d^2 \alpha \int d^2 \beta \nonumber \\
&&\exp \Bigg[ -(|\alpha |^2 +|\beta |^2 )+\lambda \alpha ^\ast \beta ^\ast
+T\alpha \beta \nonumber \\
&&\hspace{10mm} +\sqrt{T} \xi ^{\prime \ast } \alpha -\sqrt{T} \xi ' \beta -\frac{x^2 }{2}
+\frac{\xi ^{\prime 2} }{2} \Bigg] R\alpha \beta \nonumber \\
&=&\sqrt{\frac{1-\lambda ^2 }{2\pi ^5 P_{\rm det} ^{(1)} } } R
\frac{\partial }{\partial \mu ' } \int d^2 \alpha \int d^2 \beta \nonumber \\
&&\exp \Bigg[ -(|\alpha |^2 +|\beta |^2 )+\lambda \alpha ^\ast \beta ^\ast
+\mu ' \alpha \beta \nonumber \\
&&\hspace{10mm} +\sqrt{T} \xi ^{\prime \ast } \alpha -\sqrt{T} \xi ' \beta -\frac{x^2 }{2}
+\frac{\xi ^{\prime 2} }{2} \Bigg] \Bigg| _{\mu ' =T} \nonumber \\
&=&\sqrt{\frac{1-\lambda ^2 }{2\pi P_{\rm det} ^{(1)} } }
\frac{R}{T} \frac{\lambda T}{(1-\lambda T)^2 }
\left[ -\frac{\lambda T}{2(1-\lambda T)} |\xi ' |^2 +1 \right] \nonumber \\
&&
\times \exp \left[ -\frac{\lambda T}{2(1-\lambda T)} |\xi ' |^2 -\frac{(x-x_s )^2 }{2}
+\frac{\xi ^{\prime 2} }{2} \right] ,
\end{eqnarray}
where we use the relation
\begin{equation}
\hat{U}_{\rm A} (x_s ,p_s )|\Pi (x,p)\rangle _{\rm AB}
=e^{-ip_s (x-x_s )} |\Pi (x-x_s ,p-p_s )\rangle _{\rm AB} ,
\end{equation}
and
\begin{equation}
\xi ' =\frac{(x-x_s )+i(p-p_s )}{\sqrt{2} } = \xi -\xi _s .
\label{xi}
\end{equation}
The auxiliary parameter $\mu ' (=\mu +T)$ should be set to $T$
after all integration and differential operations.
From above, we obtain the homodyne probability distribution,
\begin{eqnarray}
\lefteqn{P_{\rm NG} ^{\rm (HD)}(x,p|x_s ,p_s ) } \nonumber \\
&=&\left| {_{\rm AB} \langle } \Pi (x,p)|
\hat{U} _{\rm A} (x_s ,p_s )
|\psi _{\rm NG} ^{(1)} \rangle _{\rm AB} \right| ^2 \nonumber \\
&=&\frac{1-\lambda ^2 }{2\pi P_{\rm det} ^{(1)} }
\frac{\lambda ^2 T^2 }{(1-\lambda T)^4 }
\left( \frac{R}{T} \right) ^2
\left[ -\frac{\lambda T}{2(1-\lambda T)} |\xi ' |^2 +1 \right] ^2 \nonumber \\
&&\hspace{10mm} \times
\exp \left[ -\frac{1+\lambda T}{2(1-\lambda T)} |\xi ' |^2 \right] \nonumber \\
&=&\frac{1-\lambda ^2 }{2\pi P_{\rm det} ^{(1)} }
\frac{\lambda ^2 T^2 }{(1-\lambda T)^4 }
\left( \frac{R}{T} \right) ^2 \nonumber \\
&&\left[ \left( \frac{\lambda T}{1+\lambda T} \right) ^2
\frac{\partial ^2 }{\partial \mu ^{\prime \prime 2} }
+\frac{2\lambda T}{1+\lambda T} \frac{\partial }{\partial \mu '' } +1 \right]
\nonumber \\
&&\hspace{10mm} \exp \left[ -\mu '' \frac{1+\lambda T}{2(1-\lambda T)} |\xi ' |^2 \right]
\Bigg| _{\mu '' =1} ,  \label{homodyne_pure_NG}
\end{eqnarray}
where the auxiliary parameter $\mu '' $ should be set to unity
after all differential operations.
With Eq. (\ref{homodyne_pure_NG}), the components of channel matrix can be calculated.
For example,
\begin{eqnarray}
\lefteqn{P_{\rm NG} ^{\rm (ch)} (b_{00} |a_{00} )} \nonumber \\
&=&\int _0 ^\infty dx \int _0 ^\infty dp
P_{\rm NG} ^{\rm (HD)} (x,p|\sqrt{2} \beta ,\sqrt{2} \beta ) \nonumber \\
&=&\frac{1}{4} \mathcal{D} _{\mu '' } \frac{1}{\mu '' }
\left[ 1+\textrm{erf} \left( \sqrt{\frac{1+\lambda T}{1-\lambda T} \mu '' } \beta
\right) \right] ^2
\Bigg| _{\mu '' =1} , \nonumber \\
\end{eqnarray}
where
\begin{eqnarray}
\mathcal{D} _{\mu '' } &=&\frac{(1-\lambda ^2 )\lambda ^2 T^2 }{P_{\rm det} ^{(1)}
(1-\lambda T)^2 (1-\lambda ^2 T^2 )} \left( \frac{R}{T} \right) ^2 \nonumber \\
&&\times \left[ \left( \frac{\lambda T}{1+\lambda T} \right) ^2
\frac{\partial ^2 }{\partial \mu ^{\prime \prime 2} }
+\frac{2\lambda T}{1+\lambda T} \frac{\partial }{\partial \mu '' }
+1 \right], \nonumber \\
&&
\end{eqnarray}
containing the differential operation with respect to the auxiliary parameter $\mu '' $.
Other components can be derived similarly. With these results, we can
obtain the mutual information (\ref{mutual_info_pure_NG}).

\section{Derivation of (\ref{fidelity_mixed_NG}) and (\ref{mutual_info_mixed_NG})}
\label{appendix_B}

Similar to the Appendix \ref{appendix_A}, we describe the non-Gaussian mixed state
with coherent basis,
\begin{eqnarray}
\lefteqn{ \hat{\rho }_{\rm NG} } \nonumber \\
&=&\frac{1-\lambda ^2 }{\pi ^4 \mathcal{P}_{\rm det} }
\int d^2 \alpha _1 \int d^2 \alpha _2 \int d^2 \beta _1 \int d^2 \beta _2
\nonumber \\
&&\exp \bigg[ -\frac{1+R}{2} (|\alpha _1 |^2 +|\alpha _2 |^2
+|\beta _1 |^2 +|\beta _2 | ^2 ) \nonumber \\
&&\hspace{40mm} +\lambda (\alpha _1 ^\ast \beta _1 ^\ast +\alpha _2 \beta _2 ) \bigg]
\nonumber \\
&&\times \left( e^{R(\alpha _1 \alpha _2 ^\ast +\beta _1 \beta _2 ^\ast )}
-e^{R\alpha _1 \alpha _2 ^\ast } -e^{R\beta _1 \beta _2 ^\ast } +1\right) \nonumber \\
&&\hspace{10mm} |\sqrt{T} \alpha _1 \rangle _{\rm A} \langle \sqrt{T} \alpha _2 |
\otimes |\sqrt{T} \beta _1 \rangle _{\rm B} \langle \sqrt{T} \beta _2 | \nonumber \\
&=&\sum _{i,j =0} ^1 (-1)^{i+j} \frac{1-\lambda ^2 }{\pi ^4 \mathcal{P}_{\rm det} }
\int d^2 \alpha _1 \int d^2 \alpha _2 \int d^2 \beta _1 \int d^2 \beta _2
\nonumber \\
&&\exp \bigg[ -\frac{1+R}{2} (|\alpha _1 |^2 +|\alpha _2 |^2
+|\beta _1 |^2 +|\beta _2 | ^2 ) \nonumber \\
&&\hspace{5mm} +\lambda (\alpha _1 ^\ast \beta _1 ^\ast +\alpha _2 \beta _2 )
+\gamma _i \alpha _1 \alpha _2 ^\ast +\gamma _j \beta _1 \beta _2 ^\ast \bigg] \nonumber \\
&&\hspace{10mm} |\sqrt{T} \alpha _1 \rangle _{\rm A} \langle \sqrt{T} \alpha _2 |
\otimes |\sqrt{T} \beta _1 \rangle _{\rm B} \langle \sqrt{T} \beta _2 | ,
\label{mixed_NG_coherent}
\end{eqnarray}
where $\mathcal{P}_{\rm det} $ is give by Eq. (\ref{prob_mixed_NG}), and
$\gamma _1 =R$, $\gamma _0 =0$.

With the representation (\ref{mixed_NG_coherent}), the unnormalized state after
teleportation operation is
\begin{eqnarray}
\lefteqn{ \mathcal{P} ^{\rm (tlp)} _{\rm NG} (x,p)
\hat{\rho } _{\rm NG} ^{\rm (out)} (x,p)} \nonumber \\
&=&\hat{D} _{\rm B} (\xi ) {_{\rm TA} \langle }\Pi (x,p)|
\Big[ \hat{\rho }_{\rm T} ^{\rm (in)} \otimes \hat{\rho }_{\rm AB} ^{\rm (E)} \Big]
|\Pi (x,p) \rangle _{\rm TA} \hat{D}_{\rm B} ^\dagger (\xi ) , \nonumber \\
&=&\sum _{i,j=0} ^1 (-1)^{i+j} \frac{1-\lambda ^2 }{2\pi ^3 \mathcal{P}_{\rm det} }
\int d^2 \beta _1 \int d^2 \beta _2 \nonumber \\
&&\exp \bigg[ -\frac{1+R}{2} (|\beta _1 |^2 +|\beta _2 |^2 )
+\gamma _j \beta _1 \beta _2 ^\ast +\lambda ^2 \gamma _i \beta _1 ^\ast \beta _2
\nonumber \\
&&\hspace{10mm} -\lambda \sqrt{T}Q\beta _1 ^\ast -\lambda \sqrt{T} Q^\ast \beta _2 -|Q|^2
\nonumber \\
&&\hspace{20mm} +\frac{\sqrt{T}}{2} (\xi \beta _1 ^\ast -\xi ^\ast \beta _1
+\xi ^\ast \beta _2 -\xi \beta _2 ^\ast ) \bigg] \nonumber \\
&&\hspace{30mm} |\sqrt{T}\beta _1 +\xi \rangle _{\rm B} \langle \sqrt{T} \beta _2 +\xi |,
\end{eqnarray}
where $Q$ and $\xi $ are similar to the definition in Eq. (\ref{Q}).
Therefore, the $(x,p)$ component of the fidelity is
\begin{eqnarray}
\lefteqn{\mathcal{P} _{\rm NG} ^{\rm (tlp)} (x,p)
\mathcal{F} _{\rm NG} (x,p) } \nonumber \\
&=&\sum _{i,j=0} ^1 (-1)^{i+j} \frac{1}{2\pi \mathcal{P}_{\rm det} }
\frac{1-\lambda ^2 }{1-\lambda ^2 \gamma _i \gamma _j } \nonumber \\
&&\times \exp \left[ -\mathcal{K} _{ij} \left\{
\left( x-\sqrt{2} \alpha _0 ^{\rm (r)} \right)^2
+\left( p-\sqrt{2} \alpha _0 ^{\rm (i)} \right)^2 \right\} \right], \nonumber \\
&& \label{fidelity_xp_mixed_NG}
\end{eqnarray}
where
\begin{equation}
\mathcal{K} _{ij} =\frac{1-\lambda T-\lambda ^2 \gamma _i \gamma _j
-\frac{\lambda ^2 T}{2} (\gamma _i +\gamma _j ) }{1-\lambda ^2 \gamma _i \gamma _j }
\end{equation}
and $\alpha _0 = \alpha _0 ^{\rm (r)} +i\alpha _0 ^{\rm (i)} $.

By integrating Eq. (\ref{fidelity_xp_mixed_NG}) with respect to $x$ and $p$,
we obtain the average fidelity,
\begin{eqnarray}
\lefteqn{\bar{\mathcal{F} }_{\rm NG} (\lambda )} \nonumber \\
&=&\sum _{i,j=0} ^1 (-1)^{i+j} \frac{1}{2\mathcal{P}_{\rm det} }
\frac{1-\lambda ^2 }{1-\lambda T-\lambda ^2 \gamma _i \gamma _j
-\frac{\lambda ^2 T}{2} (\gamma _i +\gamma _j ) } \nonumber \\
&= &\sum _{i,j=0} ^1 (-1)^{i+j} \mathcal{F}_{ij} (\lambda ) .
\end{eqnarray}

For the mutual information of dense coding,
we calculate the homodyne probability distribution,
\begin{eqnarray}
\lefteqn{\mathcal{P} ^{\rm (HD)} _{\rm NG} (x,p|x_s ,p_s ) } \nonumber \\
&=&{_{\rm AB} \langle } \Pi (x,p) |\hat{U} _{\rm A} (x_s ,p_s )
\hat{\rho }_{\rm NG} \hat{U} _{\rm A} ^\dagger (x_s ,p_s ) |\Pi (x,p)\rangle _{\rm AB}
\nonumber \\
&=&\frac{1}{2\pi \mathcal{P}_{\rm det} }
\frac{1-\lambda ^2 }{(1-\lambda T)^2 -\lambda ^2 \gamma _i \gamma _j } \nonumber \\
&&\times \exp \left[ -\frac{1-\lambda ^2 (T+\gamma _i )(T+\gamma _j )}
{2\{ (1-\lambda T)^2 -\lambda ^2 \gamma _i \gamma _j \} } |\xi '|^2 \right]
\nonumber \\
&= &\sum _{i,j=0} ^1 (-1)^{i+j} \mathcal{P} _{ij} (x,p|x_s ,p_s ),
\label{homodyne_mixed_NG}
\end{eqnarray}
where $\xi _s $ and $\xi '$ are given in Eq. (\ref{xi}).

With (\ref{homodyne_mixed_NG}), the component of channel matrix can be
calculated. For example,
\begin{eqnarray}
\lefteqn{\mathcal{P}^{\rm (ch)} _{\rm NG} (b_{00} |a_{00} ) } \nonumber \\
&=&\int _0 ^\infty dx \int _0 ^\infty dp
\mathcal{P} ^{\rm (ch)} _{\rm NG} (x,p|\sqrt{2}\beta  ,\sqrt{2}\beta ) \nonumber \\
&=&\sum _{i,j=0} ^1 \frac{(-1)^{i+j} }{4\mathcal{P}_{\rm det} }
\frac{1-\lambda ^2 }{1-\lambda ^2 (T+\gamma _i )(T+\gamma _j )} \nonumber \\
&&\times \left[ 1+\textrm{erf}\left( \sqrt{
\frac{1-\lambda ^2 (T+\gamma _i )(T+\gamma _j ) }
{(1-\lambda T)^2 -\lambda ^2 \gamma _i \gamma _j } } \beta \right) \right] ^2
\nonumber \\
&= &\sum _{i,j=0} ^1 (-1)^{i+j} \mathcal{C} _{ij} \left[
1+\textrm{erf}\left( \Omega _{ij} \beta \right) \right] ^2 .
\end{eqnarray}
Other components can be derived similarly. With these results,
we can obtain the mutual information (\ref{mutual_info_mixed_NG}).

\end{document}